# Multiphysics Modeling of SNAP 10A/2 Space Reactor with Cardinal

Maximiliano Dalinger[1], Elia Merzari[1], Tri Nguyen[1], Michael Seneca[1], and Richard Martineau[2]

[1] Pennsylvania State University, State College, PA
[2] Sawtooth Simulation, LLC, Idaho Falls, Idaho

Primary Author Contact Information: mgd5394@psu.edu

*The SNAP 10/A nuclear-powered satellite was launched into space in 1965. The present work discusses the development of a coupled neutronic-thermal hydraulics model with the high-fidelity multiphysics code Cardinal for SNAP 10/A. A comparison of neutronic benchmarks between OpenMC and MCNP5 shows a difference of 304±70 pcm. The primary source of difference is that MCNP5 uses a combination of cross-section libraries ENDF/B-VI and ENDF/B-V for minor isotopes. At the same time, the present work utilizes ENDF/B-VIII.0. A comparison against experimental results showed a difference of 355 pcm. A coupled model of the SNAP 10/A reactor is then developed in Cardinal. Temperature results show reasonable agreement with reference calculations.*

## I. INTRODUCTION

The interest in Space Nuclear Power has increased recently. Such is the case of the Systems for Nuclear Auxiliary Power (SNAP) reactor, developed by Atomics International, a Division of North American Aviation Inc. SNAP reactors were the subject of significant development during the 1950s and 1960s under contract to the Atomic Energy Commission. These compact nuclear reactors were designed for remote or automated operation to generate electrical power for Earth satellites and other space vehicles. The SNAP 10/A was the first nuclear-powered satellite, launched into space in 1965 and capable of generating an electrical power of 500 W (Ref. 1, Ref. 2). The reactor stopped working after 43 days because of a non-nuclear electrical failure.

In the present work, we propose to develop a coupled neutronic-thermal hydraulics model of the SNAP 10/A reactor using the high-fidelity multiphysics code Cardinal. This paper is organized as follows: Section II introduces the SNAP 10/A space reactor and its characteristics. Section III briefly introduces the computational tools used for the simulations. Section IV details the neutronic benchmarking performed with the neutronic code OpenMC. Section V explains the coupled neutronic-thermal hydraulic model developed in Cardinal and presents its results. Finally, Section VI provides paper conclusions and discusses future work.

## II. SNAP 10/A SPACE REACTOR

The SNAP 10/A reactor core has 37 moderator-fuel elements arranged in a hexagonal array[1,2]. Fuels are composed of hydride zirconium-uranium alloy with 10 wt. % uranium (enrichment of 93 wt. % $^{235}$U). The cladding is composed of Hastelloy N, surrounding the fuel radially, and closed by solid plugs at both ends. The internal surface of the cladding tube is thinly coated with a small amount of samarium oxide to prevent the hydrogen from diffusing out of the element at operating temperatures. The core assembly is contained within a cylindrical vessel of 316 stainless steel. Six internal side reflectors of beryllium are used to round out the hexagonal core configuration and fill the void spaces in the core vessel. External beryllium reflectors surround the vessel, composed of a static attached part and four control drums. The core is cooled by NaK-78 sodium-potassium alloy[3]. Figure 1 presents a reactor diagram. Table I summarizes principal geometrical parameters, and Table II principal reactor design parameters. The reactor employs thermoelectric elements to convert fission heat into electrical energy. Electrical power is generated by the heat transfer through the elements with heat rejection to space.

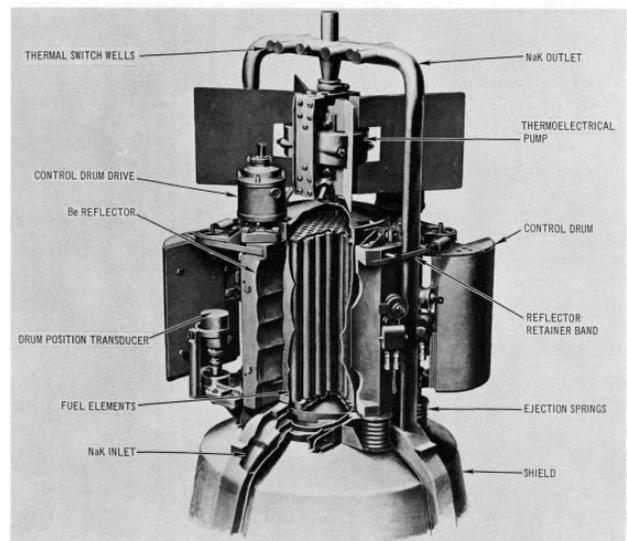

**Fig. 1.** SNAP 10/A reactor diagram[2].

**TABLE I.** Principal geometric parameters of SNAP 10/A reactor core.

| Parameter | Value |
|---|---|
| Fuel radius [cm] | 1.53924 |
| Coating radius [cm] | 1.56210 |
| Cladding radius [cm] | 1.58750 |
| Pitch [cm] | 3.20040 |
| Active length [cm] | 31.0515 |
| Fuel rod length [cm] | 31.623 |
| Number of fuel elements | 37 |
| Vessel internal radius [cm] | 11.27125 |
| Vessel external radius [cm] | 11.34999 |

**TABLE II.** Principal core design parameters of SNAP 10/A reactor core.

| Parameter | Value |
|---|---|
| Reactor power [kW] | 34.0 |
| Inlet temperature [K] | 755.37 |
| Outlet temperature [K] | 816.48 |
| Average temperature [K] | 783.15 |
| Mass flow rate [kg/s] | 0.6199 |
| NaK density [kg/m$^3$] | 755.92 |
| NaK viscosity [Pa*s] | 1.8835e-4 |
| NaK heat capacity [J/kg*K] | 879.903 |
| NaK thermal conductivity [W/m*K] | 26.2345 |
| Fuel thermal conductivity [W/m*K] | 22.484 |
| Coating thermal conductivity [W/m*K] | 1.729 |
| Cladding thermal conductivity [W/m*K] | 18.852 |

## III. COMPUTATIONAL TOOLS

### III.A. OpenMC

OpenMC[4] is an open-source high-fidelity Monte Carlo neutron and photon transport simulation code. It can perform cell tallies and *k*-eigenvalue calculations, considering Doppler broadening of cross-sections and many other features.

### III.B. MOOSE

MOOSE[5] (Multiphysics Object-Oriented Simulation Environment) is a finite element framework for solving fully coupled, fully implicit multiphysics simulations. It executes multiple sub-applications simultaneously and transfers data between the scales. It discretizes the space using the PETSc non-linear solver and libMesh. Some of its capabilities include automatic differentiation, scaling to a large number of processors, hybrid parallelism, and mesh adaptivity. The Idaho National Laboratory developed MOOSE.

### III.C. NekRS

NekRS[6] is an open-source Spectral Element Method (SEM) Computational Fluid Dynamics (CFD) code to simulate fluid dynamics and heat transfer. Its capabilities include RANS modeling, Large Eddy Simulation (LES), and Direct Numerical Simulation (DNS). It supports both CPU and GPU backends.

### III.D. Cardinal

Cardinal[7] is an open-source application that wraps OpenMC and NekRS codes within the MOOSE framework. It uses the MOOSE data transfer implementation to perform high-fidelity coupled neutronic-thermal hydraulics calculations. It can also be coupled with any MOOSE application, enabling a broad set of multiphysics capabilities.

## IV. NEUTRONICS BENCHMARKING

Ref. 1 presents computational benchmarks for several experimental configurations of the SNAP 10A/2-type reactor. These experiments considered varied water immersion and reflection conditions, fuels, absorber sleeves, and other factors. A total of 73 different experimental configurations were described, and all were modeled in MCNP5. The first part of this project was reproducing these neutronic benchmarks in OpenMC.

All 73 experimental configurations were modeled in OpenMC using Constructive Solid Geometry (CSG). Figure 2 presents an example of the OpenMC model for configuration "fig12", where blue is water, yellow is beryllium, pink is for Lucite rods, black is for stainless steel, white is void, and fuels in different colors represent different burnups. This case was the only experimental configuration with the beryllium external reflector. For all cases, simulations used 500000 particles, with 10 inactive batches and 200 total batches, which resulted in uncertainties below 12.5 pcm for the effective multiplication factor *Keff*. The cross-section library used was ENDF/B-VIII.0.

Figure 3 presents the *Keff* comparison of OpenMC with MCNP5. As can be seen, there is a good agreement between both codes. The average absolute difference was 304±70 pcm, and the maximum difference was -725±71 pcm for case "8490l3". The principal source of difference is the cross-section library. MCNP5 simulations used a combination of ENDF/B-VI and ENDF/B-V for minor isotopes. Figure 4 compares OpenMC *Keff* with experimental results. In this case, only 58 experiments reported a value for the *Keff*, and no uncertainty was reported. The average absolute difference was 355 pcm, and the maximum difference was -1048 pcm for case "8490e7". Both comparisons show a good agreement between OpenMC and MCNP5 and experimental results. Therefore, it is considered that OpenMC can be applied to perform neutronics calculations for the SNAP 10A/2 reactor core.

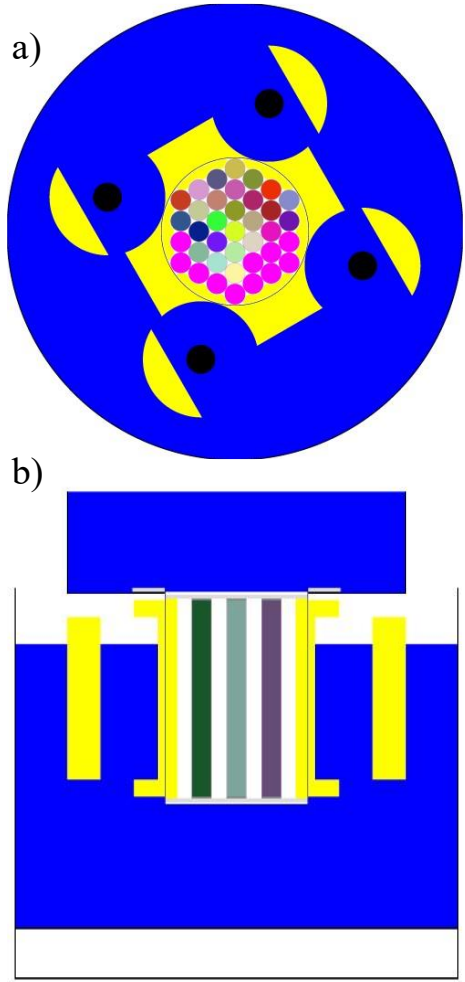

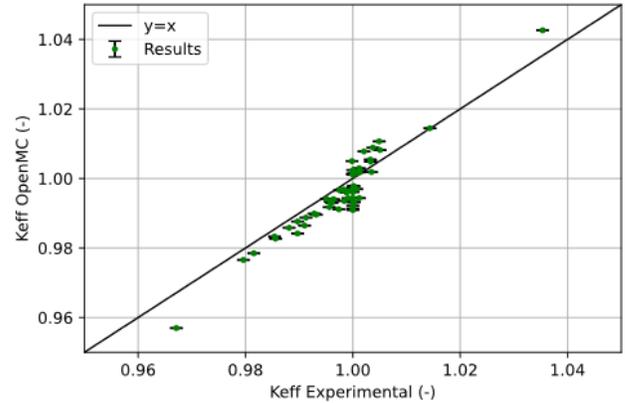

**Fig. 4.** *Keff* comparison between OpenMC and experimental results.

## V. COUPLED MODEL

### V.A. OpenMC Model

Neutronics equations are solved in OpenMC to obtain the heat source distribution. The SNAP 10/A core model for coupling calculations is derived from case "fig12" from Ref. 1. NaK replaces water for the coolant, and all water tanks around the core are removed. The core is composed of 37 fresh SCA-4 fuels. Both internal and external beryllium reflectors are present. The position of the control drums is selected to get a critical multiplication factor. Simulations used 300000 particles, with 10 inactive batches and 50 total batches, obtained from a Shannon Entropy analysis.

Fuel cells are discretized axially and radially to account for temperature variation. A meshing analysis determined that the optimal discretization consists of 20 axial, 4 radial, and 6 azimuthal cells per fuel unit and 20 axial cells for each coolant channel, resulting in a total of 18723 cells for the model. Temperature feedback from the MOOSE heat conduction model is applied to the fuel and coolant, while density feedback is applied to the coolant exclusively. Figure 5 presents the OpenMC model for coupling calculations, where red is NaK, yellow is beryllium, dark gray is Hastelloy N, black is stainless steel, and violet is the coating between fuel and cladding. It also presents the CSG OpenMC geometry, where each color represents an individual cell. The reactor geometry is surrounded by a void cell with vacuum boundary conditions.

**Fig. 2.** OpenMC model for case "fig12". a) Superior view, b) lateral view.

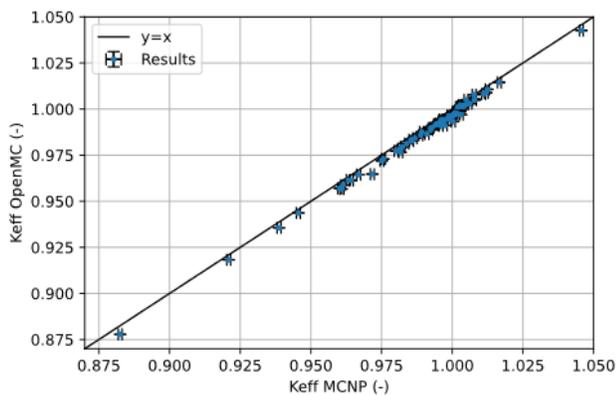

**Fig. 3.** *Keff* comparison between MCNP5 and OpenMC.

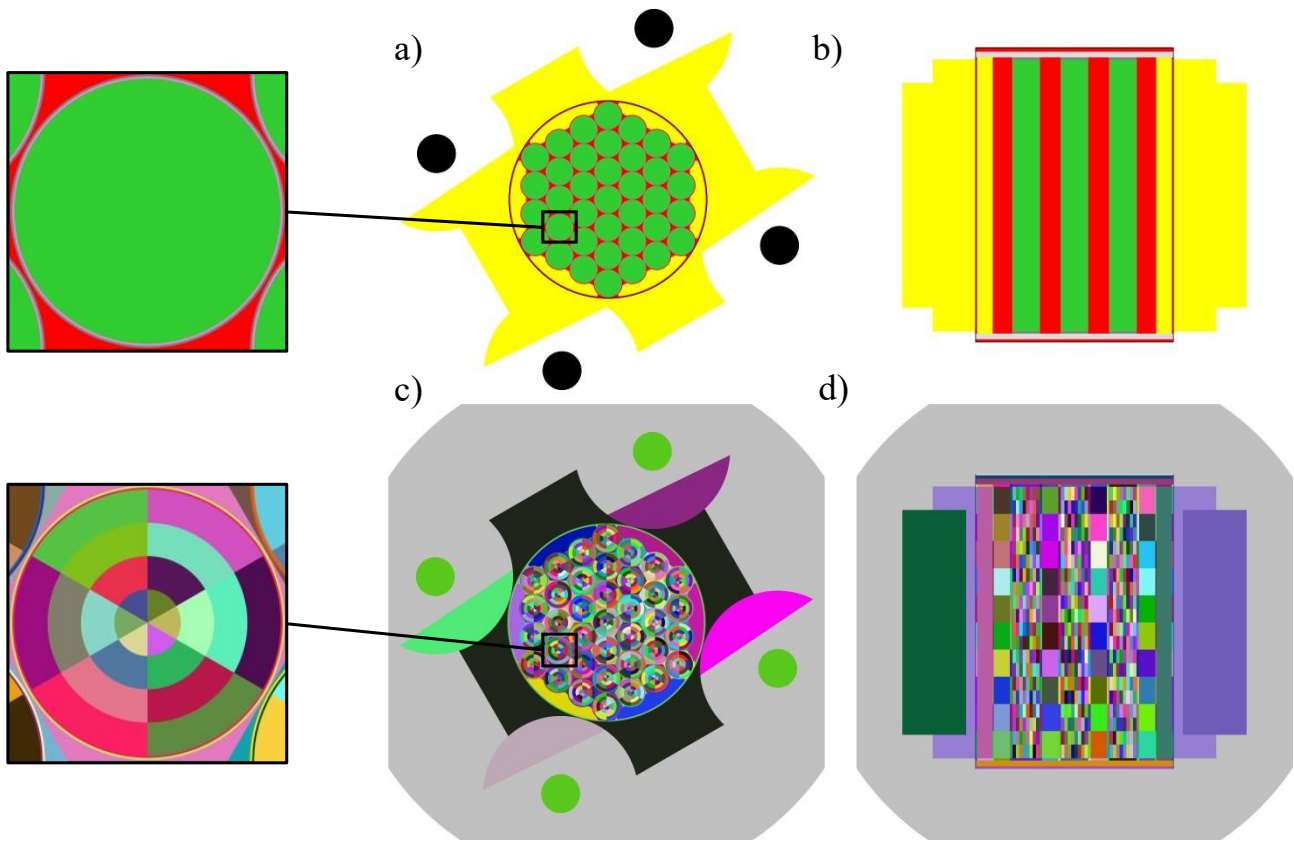

**Fig. 5.** OpenMC model for coupling calculations. a) Superior view colored by material, and b) lateral view colored by material, c) superior view colored by cell, and d) lateral view colored by cell.

**V.B. MOOSE Heat Conduction Model**

MOOSE has an extensive library of physics modules, including the Heat Transfer Module. In this work, we use it to solve heat conduction equations and calculate the temperature distribution within fuel rods. The heat source is obtained with OpenMC, and the surface temperature of the rods, which corresponds to the fluid temperature in contact with the solid, is obtained with NekRS. Adiabatic boundary conditions were considered for the vessel and beryllium reflector surfaces, and the top and bottom of the fuel.

This geometry is also used to transfer the coolant temperature from NekRS to OpenMC through volume averaging. It also calculates the coolant density[3]. Fluid temperature and density are not considered when solving heat conduction in the solid; only the cladding surface temperature is considered.

Figure 6 presents the mesh used for heat conduction. This model considers the fuel, coating, and cladding for the active length. The geometry was generated using MOOSE's Reactor Module. The mesh used for the calculation has 40 axial, 4 radial, and 60 azimuthal cells are required.

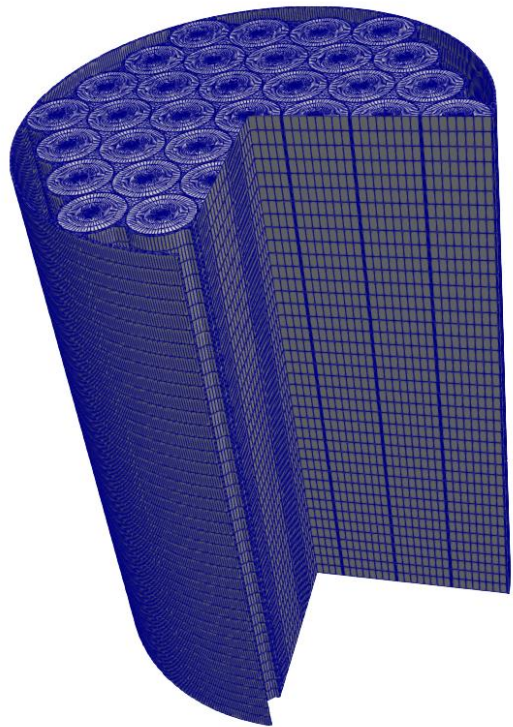

**Fig. 6.** MOOSE model for heat conduction in solid.

**V.C. NekRS Model**

In the present work, NekRS solves fluid mass, momentum, and energy conservation equations via LES. It takes the heat flux calculated in MOOSE and uses it as a boundary condition. The remaining boundary conditions are adiabatic for lateral surfaces corresponding to the vessel and beryllium reflectors, and uniform inlet temperature 755.37 K. The bottom of the model is extended to generate a non-uniform velocity profile considering a recycling velocity just below the active region. Adiabatic boundary conditions are imposed on the walls of the extended region, so only heat transfer to the fluid is allowed in the active region. Constant NaK thermal properties are considered. For the conditions in Table II, the Reynolds number is 2357.05, and the Prandtl number is 6.317e-3.

Figure 7 shows a three-dimensional mesh generated with the software Gmsh to represent the fluid. It was created using approximately $1.252*10^6$ hexahedral elements. A polynomial order of 5 was selected, resulting in about $2.706*10^8$ Gauss–Lobatto-Legendre (GLL) quadrature points. After a sensitivity study, the resolution was found adequate for the problem.

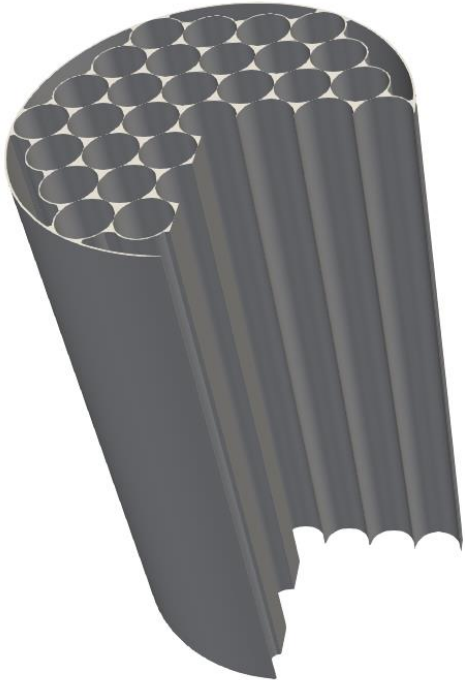

**Fig. 7.** NekRS model for the fluid.

**V.D. Cardinal Model**

Figure 8 presents the coupling procedure used in Cardinal. OpenMC runs a k-eigenvalue calculation of one time-step of $MN\Delta t_{nek}$ using an initial uniform solid temperature. It calculates the heat source $q'''$, which is sent to MOOSE. The heat conduction equation is solved in MOOSE for a time step of $N\Delta t_{nek}$. MOOSE uses the heat source and an initial linear wall temperature in the cladding surface to calculate the solid temperature $T_{solid}$ and heat flux in the cladding surface $q''$. The heat flux is sent to NekRS, which performs $N$ calculations of time step $\Delta t_{nek}$ to calculate the fluid velocity and temperature. Then NekRS sends the wall temperature $T_{wall}$ and coolant temperature $T_{coolant}$ to MOOSE. Subsequently, MOOSE calculates the coolant density $\rho_{density}$ with the coolant temperature and sends them with the fuel temperature $T_{fuel}$ to OpenMC. The process is repeated until the fluid temperature and velocity, fuel temperature, heat source distribution, and multiplication factor are converged. For the present work, $M = 5$, $N = 500$, and $\Delta t_{nek} = 4.0 * 10^{-3}$. Therefore, the OpenMC model receives feedback for the fuel temperature, coolant temperature and density. The temperature for all remaining material is 783.1 K, which corresponds to the average fluid temperature from Ref. 2.

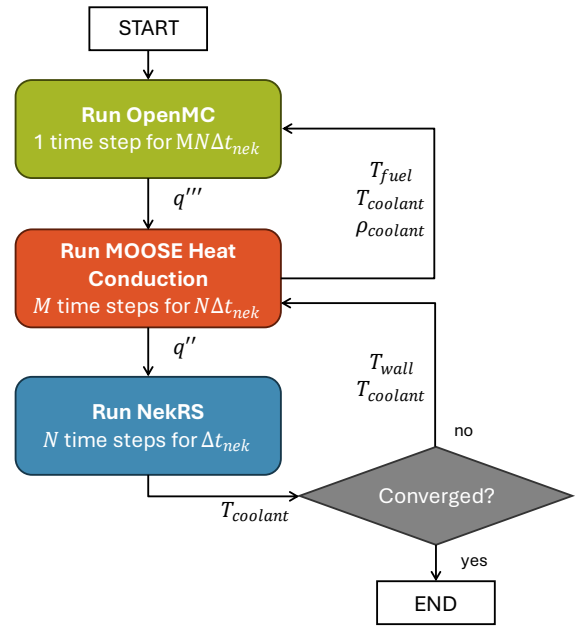

**Fig. 8.** Cardinal coupling procedure.

**V.E. Results**

Simulations were performed on the supercomputer Frontier using GPUs for NekRS and CPUs for OpenMC and MOOSE. The multiplication factor obtained is 1.00086 ± 0.00024, and the peaking factor is 1.98. Figure 9 presents the results for the fission heat source obtained in OpenMC. The heat source is approximately sinusoidal in the axial direction, with a slight shift towards the core inlet because of reactivity feedback effects produced by coolant and fuel temperature. Radially, the heat source maximum is in the center of the core and decreases towards the periphery fuels. However, fuel elements close to beryllium reflectors produce more heat than those in the corners because of increased neutron reflections.

Figure 10 compares the midplane fuel temperature obtained in MOOSE using the Heat Conduction module and coolant temperature obtained in NekRS with OpenMC cell temperature. Cell temperature is calculated as volume averages of all MOOSE mesh elements that map on each OpenMC cell. Here, it is possible to see how fuels close to beryllium reflectors have higher temperatures than fuels located on the corners, produced by the heat source distribution.

Figure 11 presents the temperature results for the solid obtained in MOOSE and the fluid from NekRS. Figure 12 depicts the fluid velocity obtained in NekRS. As can be seen, the fluid presents turbulence in the regions where the gap between fuels is wide, which increases the heat exchange and causes the fuel temperature close to these regions to decrease. Additionally, stagnation points in the fluid velocity appear where the gap is narrow between fuels, which causes the fluid temperature to increase in these regions. Figure 13 shows the coolant's cell density used in OpenMC calculations. As can be seen, it decreases inside the core, corresponding to an increment of the temperature. However, since the periphery region does not receive heat, the coolant temperature in this region is invariant, and so does the coolant density.

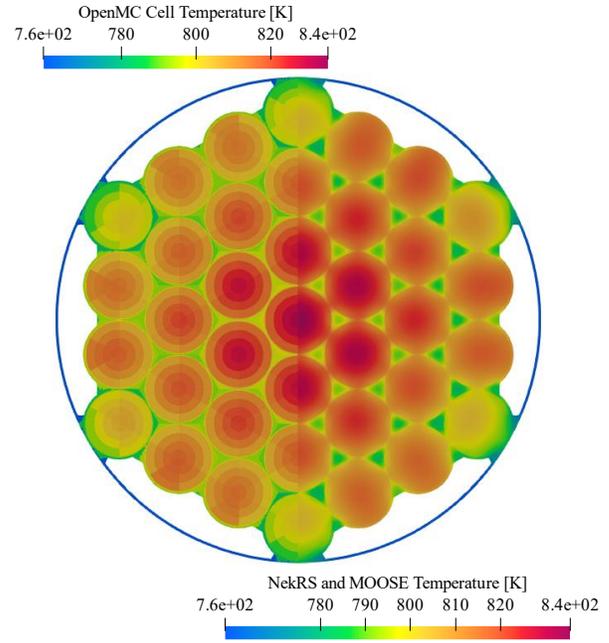

**Fig. 10**. Solid temperature comparison between OpenMC and MOOSE at the midplane.

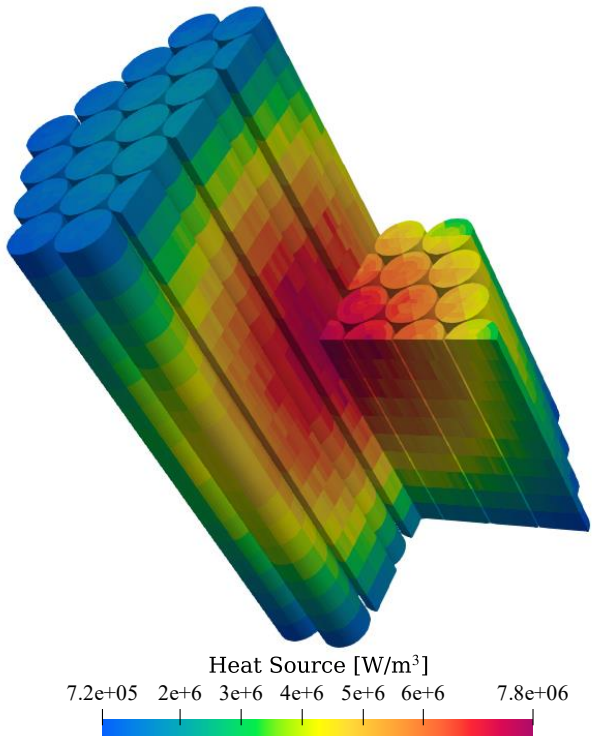

**Fig. 9.** Heat source distribution obtained in OpenMC.

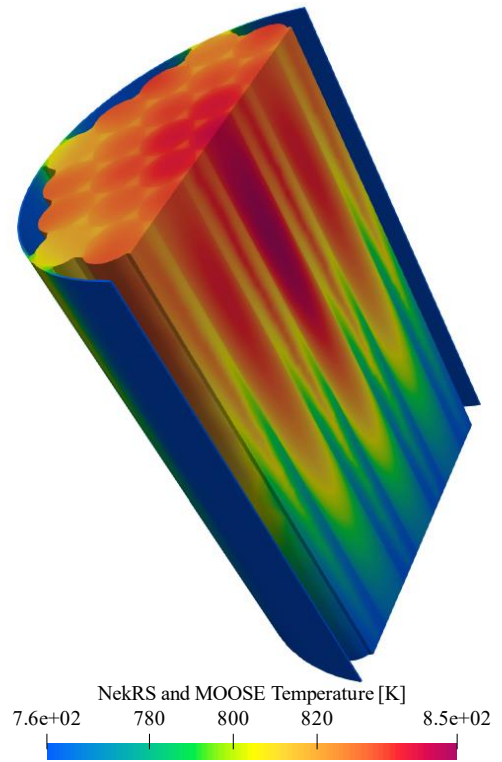

**Fig. 11.** Solid and fluid temperature obtained in MOOSE and NekRS respectively.

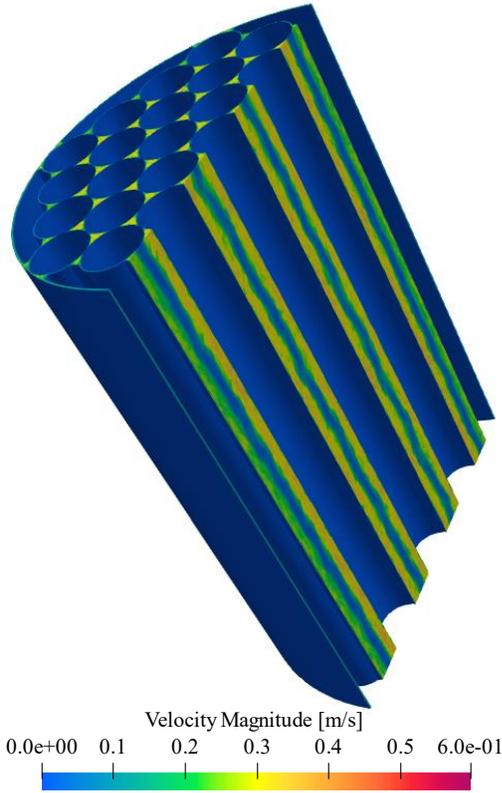

**Fig. 12.** Fluid velocity obtained in NekRS.

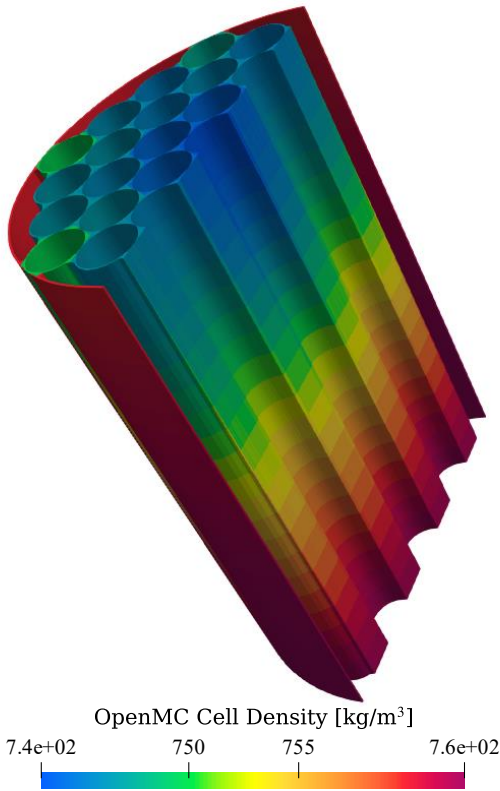

**Fig. 13.** Coolant cell density used in OpenMC.

Figure 14 presents the results of the average temperature versus axial position for the fuel centerline and fluid. This figure also compares with results from Ref. 2. We can see differences from reference results. The first source of difference is that we used adiabatic boundary conditions for the fuel rod top and bottom, while in Ref. 2, the temperature was fixed. The second source of difference is that the reference considered heat transfer with beryllium reflectors, which was not considered in our model. The third source of difference is that the reference used a sinusoidal heat source distribution, while the distribution obtained in this work is slightly moved towards the core bottom. Additionally, the reference model is 1D, while our model is 3D. For the fuel, the average temperature is 805.6 K, and the maximum is 853.3 K, a difference of -0.9 K from the result obtained in Ref. 2. For the fluid, the maximum is 843.2 K, and the average is 786.3 K, 3.0 K higher than what is reported in Ref. 2 as expected by the differences outlined. The fluid is expected to receive more heat than the reference. Additional calculations will be performed to assess the 3D model proposed here more comprehensively.

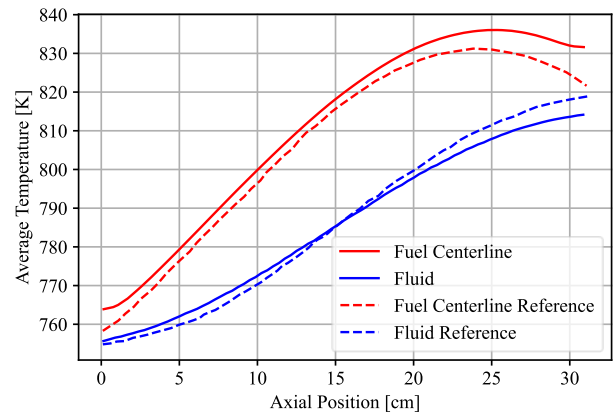

**Fig. 14.** Results for average temperature distributions for fuel, coating, cladding and fluid, compared with results from Ref. 2.

## VI. CONCLUSIONS

Neutronics benchmarks for the SNAP 10/A space reactor were performed in OpenMC. Results showed an average absolute difference of 304±70 pcm when compared with MCNP5, and 355 pcm with experimental results. The main source of difference with MCNP5 is attributed to the use of different cross-section libraries. MCNP5 simulations used a combination of ENDF/B-VI and ENDF/B-V for minor isotopes, while the present work utilized ENDF/B-VIII.0 library. The agreement was judged to be excellent with the experiment.

Coupled neutronics-thermal hydraulics calculations were performed for the SNAP 10/A space reactor with the code Cardinal. OpenMC was used to solve neutronic equations, NekRS to solve energy, momentum, and mass equations in the fluid, and the MOOSE's Heat Conduction module to calculate temperatures in the fuel rods. The model considered fuel temperature and coolant temperature and density feedback. Results showed reasonable behavior for the heat source, temperature, coolant density and velocity. Low-velocity regions between fuel rods with narrow gaps produce a high temperature in the fluid. Fuels close to the beryllium reflectors produce higher power than those on the corners of the hexagonal array. This power increase is because of an increase in the neutron flux and, therefore, the heat source. These effects are expected and lead to an overall reasonable agreement with existing sources.

For future work, we plan to improve further the modeling presented here by including additional effects that are expected to improve accuracy. Additional sensitivity analysis will be performed, providing confidence in using Cardinal to model SNAP-10A.

**ACKNOWLEDGMENTS**

We acknowledge Victor Coppo Leite for his efforts during the initial stage of the development of the NekRS model.